\documentclass[10pt,conference,letterpaper]{IEEEtran}
\IEEEoverridecommandlockouts
\usepackage{cite}
\usepackage{amsmath,amssymb,amsfonts}
\usepackage{algorithmic}
\usepackage{multicol}
\usepackage{graphicx}
\usepackage{textcomp}
\usepackage{xcolor}
\usepackage{xcolor}
\usepackage{todonotes}
\usepackage{enumitem}
\usepackage{microtype}
\usepackage{url}
\usepackage{balance}
\usepackage{longtable}
\usepackage{ wasysym }
\usepackage[style=iso]{datetime2}

\def\BibTeX{{\rm B\kern-.05em{\sc i\kern-.025em b}\kern-.08em
    T\kern-.1667em\lower.7ex\hbox{E}\kern-.125emX}}

\begin{document}

\title{Hardening X.509 Certificate Issuance using Distributed Ledger Technology
\thanks{This work has been supported by the German Federal Ministry of Education and Research, project VITAF, grant 16KIS0834 and the German-French Academy for the Industry of the Future.
\protect\\
\indent Author's version -- Final paper to appear in 2020 IEEE/IFIP International Workshop on Security for Emerging Distributed Network Technologies (DISSECT) co-located with the IEEE/IFIP Network Operations and Management Symposium (NOMS) 2020, Budapest, Hungary.}
}

\author{\IEEEauthorblockN{Holger Kinkelin, Richard von Seck, Christoph Rudolf and Georg Carle}
\IEEEauthorblockA{Technische Universit\"at M\"unchen, Department of Informatics,
Chair of Network Architectures and Services \\
85748 Garching bei M\"unchen, Germany\\ \\
\{kinkelin, seck, rudolfc, carle\}@net.in.tum.de}}

\maketitle

\begin{abstract}
The security of cryptographic communication protocols that use X.509 certificates depends on the correctness of those certificates.
This paper proposes a system that helps to ensure the correct operation of an X.509 certification authority and its registration authorities.
We achieve this goal by enforcing a policy-defined, multi-party validation and authorization workflow of certificate signing requests.
Besides, our system offers full accountability for this workflow for forensic purposes.
As a foundation for our implementation, we leverage the distributed ledger and smart contract framework Hyperledger Fabric.
Our implementation inherits the strong tamper-resistance of Fabric which strengthens the integrity of the computer processes that enforce the validation and authorization of the certificate signing request, and of the metadata collected during certificate issuance.
\end{abstract}


\section{Motivation}
\label{sec:intro}

\emph{X.509}~\cite{x509} is a widely used standard that specifies a public key infrastructure and the format of public key certificates.
X.509 certificates are issued by \emph{Certificate Authorities (CA)} and bind the identity of an entity to the public part of an asymmetric key pair owned by that entity.
An example of such a binding is the link between a Web server's URL and its public key, or between a person's name, e-mail address, and her public key.
To prove authenticity, certificates are digitally signed by a CA.

X.509 certificates are used in a variety of cryptographic communication protocols and help to achieve the security goals of confidentiality, integrity, and authenticity.
For this reason, X.509 must be regarded as a cornerstone of network security.
However, the security of communication protocols that use certificates depends on the correctness of those certificates.
Unfortunately, situations can occur that lead to authentic certificates that are either incorrect (due to a mistake) or even fraudulent (due to an attack).
In both cases, the above security goals are harmed as the identity claimed in such a certificate and the public key it contains do \emph{not} belong to the same entity.

The attack vectors that can lead to incorrect/fraudulent certificates include \emph{Registration Authorities (RA)} who validate \emph{Certificate Signing Requests (CSR)} on behalf of a CA.
This process typically involves a human RA member who is prone to mistakes or may even act maliciously.
Furthermore, RA members use applications running on computers to communicate with the CA that can be compromised if not properly secured~\cite{racomp}.
Finally, CAs themselves might be improperly secured, risking compromise of the CA's signing key that leads to extensive abuse~\cite{digicomp}.

In this paper, we focus on improving the correct operation of an X.509 CA and its RAs.
In this context, "correct" means that only correctly validated certificates are issued by the CA.
We propose a solution that provides both increased robustness of the certificate issuance process and accountability for forensic purposes.
Forensic analyses are, for instance, helpful to identify compromised or malicious RA members.
The certification process is improved by enforcing a policy-defined, multi-party validation and authorization workflow of CSRs that involves more than just one RA member.
Our system was designed to run on top of the distributed ledger and smart contract framework Hyperledger Fabric \cite{Androulaki2018}.
This framework provides us strong tamper-resistance concerning the integrity of computer processes that enforce the CSR validation and authorization, and of collected accounting information.

In this work, we will mainly refer to corporate CAs that issue S/MIME certificates for e-mail communication~\cite{rfc5751}.
However, our findings can be applied to other scenarios.

The rest of this paper is structured as follows:
In Section~\ref{sec:back} we provide background information on X.509 certificate issuance and Hyperledger Fabric.
Next, we analyze and detail security problems related to CAs and RAs, and define requirements for a suitable solution in Section~\ref{sec:analysis}.
The design and implementation of our system are described in Section~\ref{sec:design}.
We discuss our work in Section~\ref{sec:discuss} and compare it to related work in Section~\ref{sec:rw}.
The paper is concluded in Section~\ref{sec:conclusion}.
 \section{Background}
\label{sec:back}

This section provides information on X.509 certificate issuance and the Hyperledger Fabric framework.

\subsection{Validation Process of an X.509 Certificate Signing Request}
\label{sec:back:pki}

\emph{Certificate Authorities (CA)} impose strict rules on their \emph{Registration Authorities (RA)} that describe how \emph{certificate signing requests (CSR)} must be validated before the certificate can be signed.
As an example, we describe the validation process of a CSR for an S/MIME certificate imposed by the CA of the German National Research and Education Network (\emph{DFN-PKI}~\cite{dfnpki}).
The DFN-PKI is frequently used by German universities and other research institutions, and is a sub-CA of the Deutsche Telekom CA.

As a first step, the \emph{certificate requester} generates a new asymmetric key pair and a CSR, which contains the public key, the requester's name, and e-mail address.
The CSR is then transmitted to the CA.

The requester must now meet in person with one member of her university's or department's RA who then validates the CSR.
The RA member checks the requester's identity using an official identity document and whether the requester is the owner of the claimed e-mail address.
The RA member fills out a paper form documenting the validation process and notes the last digits of the serial number of the submitted identification document to prove that he or she has verified the identification document.
Finally, the paper form is signed by the certificate requester and RA member, and filed by the RA.

After having validated the CSR, the RA member authorizes it using an application installed on her computer.
The CA now issues the new certificate and delivers it by e-mail to the certificate requester.

\subsection{Hyperledger Fabric}
\label{sec:back:hlf}

\emph{Hyperledger Fabric}~\cite{Androulaki2018} is a private and permissioned \emph{distributed ledger} and \emph{smart contract} framework developed by the Hyperledger project of the Linux Foundation~\cite{hyperpro}.

A \emph{distributed ledger} is ``a type of database that is spread across multiple sites''.
Its ``records are stored one after the other in a continuous ledger'' and ``can only be added when the par\-ti\-ci\-pants reach a quorum''~\cite[p. 17]{dlt}.
For these reasons, distributed ledgers are append-only and provide improved protection of data integrity and availability compared to centralized approaches.

\emph{Chaincode} is a concept of Fabric comparable to \emph{smart contracts} known from public blockchains such as Ethereum~\cite{ethereum2018}.
It can be used to implement any business logic and its unique feature is that multiple instances on different nodes of a Fabric network.
Chaincode can be executed by Fabric clients by sending \emph{transactions} into the Fabric network.
In this way, clients can manipulate the state of the distributed ledger, e.g., create a new data structure representing an asset or create a modified version (an update) of an existing one.
However, only if a certain, policy-defined number of chaincode instances endorse the transaction, the state of the distributed ledger will change.
Through this transaction approval model, Fabric offers Byzantine fault-tolerant execution of processes.
 \section{Problem Analysis and Requirements}
\label{sec:analysis}

This section defines attack vectors on X.509 certificate issuance and defines requirements on a solution able to mitigate these.

\subsection{Attack Vectors}
\label{sec:analysis:threat}

As a first step, we want to exemplarily analyze the X.509 certificate issuance workflow of the DFN-PKI, introduced in Section~\ref{sec:back:pki}, for possible attack vectors (AV).

\textbf{AV1: Compromising a CA:}
The CA, as a black box entity, was manipulated in a way that results in the issuance of an authentic certificate with fraudulent content.
For instance, the attacker obtained access to the CA's private key and can sign fraudulent certificates without any further authorization.~\cite{digicomp}

\textbf{AV2: Compromising an RA Member:}
The technical infrastructure used by an honest RA member, such as his client computer, was manipulated in a way that leads to the authorization of an incorrect or fraudulent CSR.~\cite{racomp}

\textbf{AV3: Negligent RA Member:}
An error of an honest RA member results in the unintended authorization of an incorrect or fraudulent CSR.

\textbf{AV4: Malicious RA Member:}
A dishonest RA member intentionally authorizes a fraudulent CSR.

\subsection{Requirements}
\label{sec:analysis:reqs}

The goal of this work is to prevent that the attack vectors defined in Section~\ref{sec:analysis:threat} can be exploited to issue incorrect or fraudulent certificates.
In addition, we want to collect metadata for forensic analyses that, for instance, allow the identification of malicious RA members.
To achieve this goal, our solution must meet the following requirements (R):

\textbf{R1: Multi-party CSR Validation:}
The validation of a CSR must not depend on only one RA member.
Instead, the validation of a CSR must be performed by multiple parties, i.e. two or more RA members.

\textbf{R2: Tamper-Resistant Enforcement of the Certificate Issuance Workflow:}
Computer processes that enforce the validation workflow of a CSR and finally authorize certificate issuance must be tamper-resistance.

\textbf{R3: Accountability of CSR Validation:}
The validation workflow of a CSR (validating RA member's identity, verified target identity, etc, ...) must be logged transparently.

\textbf{R4: Accountability of Certificate Issuance:}
When a certificate is signed by a CA, its issuance must be logged transparently.

\textbf{R5: Tamper-Resistance of Accountability Information:}
Metadata collected during CSR validation and certificate issuance must be protected concerning data integrity and availability.
 \section{Design and Implementation}
\label{sec:design}

This section gives an overview of our system and describes technically the interaction between users, system and CA.

\subsection{Overview}

\begin{figure*}[tb]
  \centering
    \includegraphics[width=0.7\textwidth]{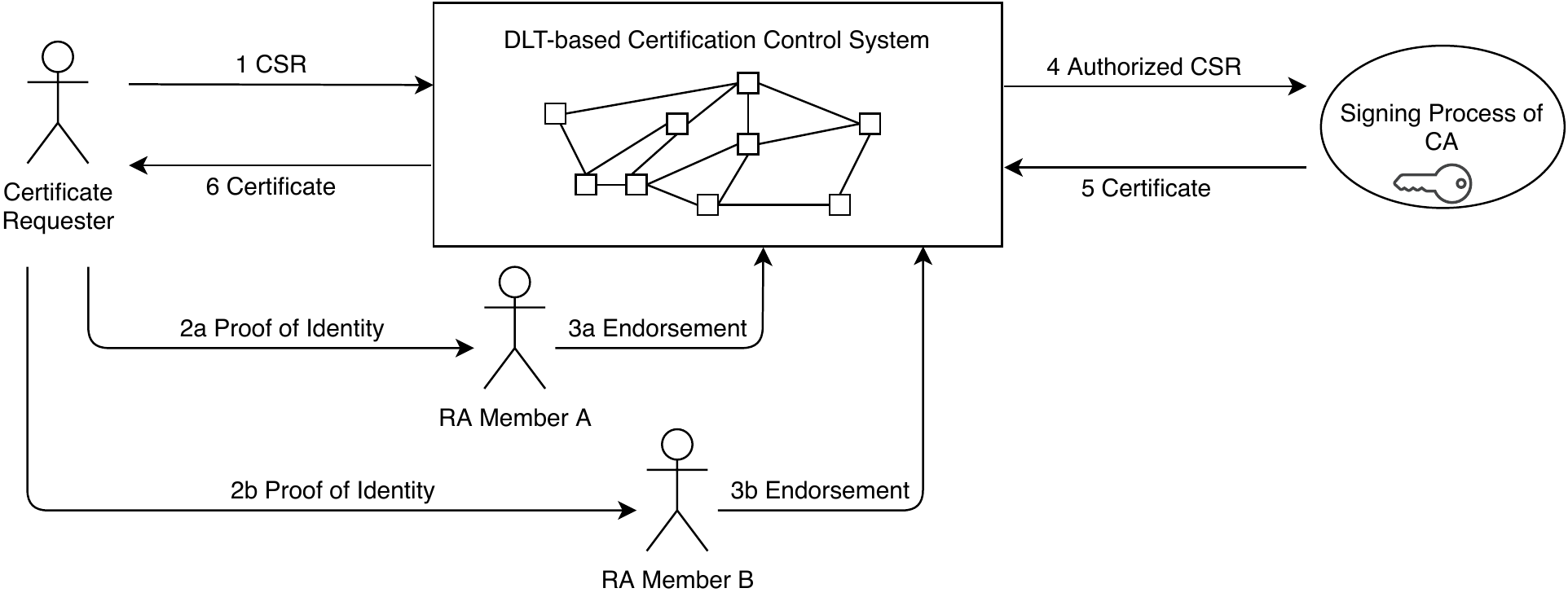}
\caption{System overview and interactions between entities}
  \label{overview}
\end{figure*}

Figure~\ref{overview} depicts an overview of our system and the interactions between entities, which follows the certificate issuance process explained in Section~\ref{sec:back:pki}.

The core of our system is the distributed ledger-based \emph{Certification Control System (CCS)}, which orchestrates and controls the certificate issuance process.
The CCS is equipped with a \emph{Certification Policy} that expresses conditions that must be met before a CSR is regarded as being authorized as well as permissions of RA members.

Permissions specify which RA members are allowed to validate and endorse specific CSRs.
For instance, RA1 is responsible for CSRs belonging to subdomain x, RA2 is responsible for subdomain y.
This prevents an RA member from endorsing a CSR of a domain for which she is not responsible.

Conditions express, for example, how many (and which) RA members must have individually endorsed a CSR before it is finally considered to be authorized. This is useful to enforce a more thorough validation of CSRs that, for instance, belong to certificate requesters with stronger security demands.

Throughout the entire certificate issuance process (handing in the CSR, its multi-party validation, its authorization, and signing the certificate), metadata is collected and persisted in the distributed ledger underlying the CCS, making the entire process fully accountable.

\subsection{Implementation}

\begin{figure}[bp]
  \centering
    \includegraphics[width=\columnwidth]{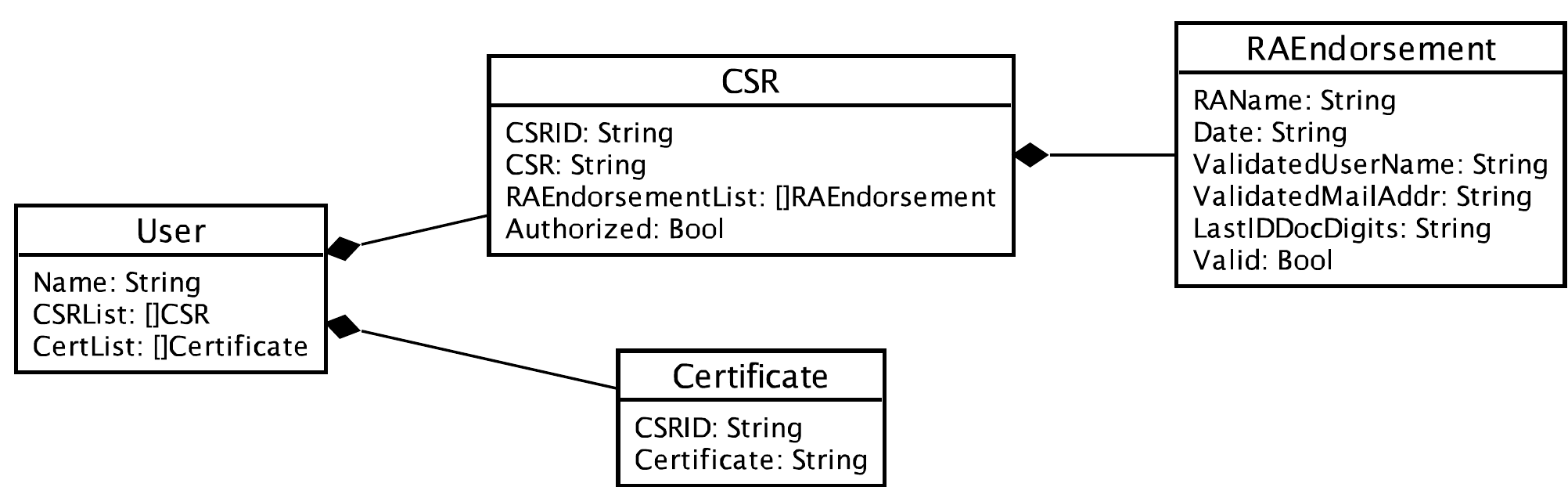}
  \caption{User-centered data model of the certificate issuance process (simplified)}
  \label{data}
\end{figure}

We have implemented our CCS in chaincode that runs on nodes of a Hyperledger Fabric network.
This chaincode creates and updates various data structures stored in the distributed ledger that model users, CSRs and certificates, see Figure~\ref{data}, RA members, the certification policy and other concepts.
The chain code is also used to determine whether a specific CSR can be considered authorized.
We will now detail how our system uses chaincode to process user input and enforce a policy-defined CSR validation process.

\begin{figure}[h]
  \centering
    \includegraphics[width=\columnwidth]{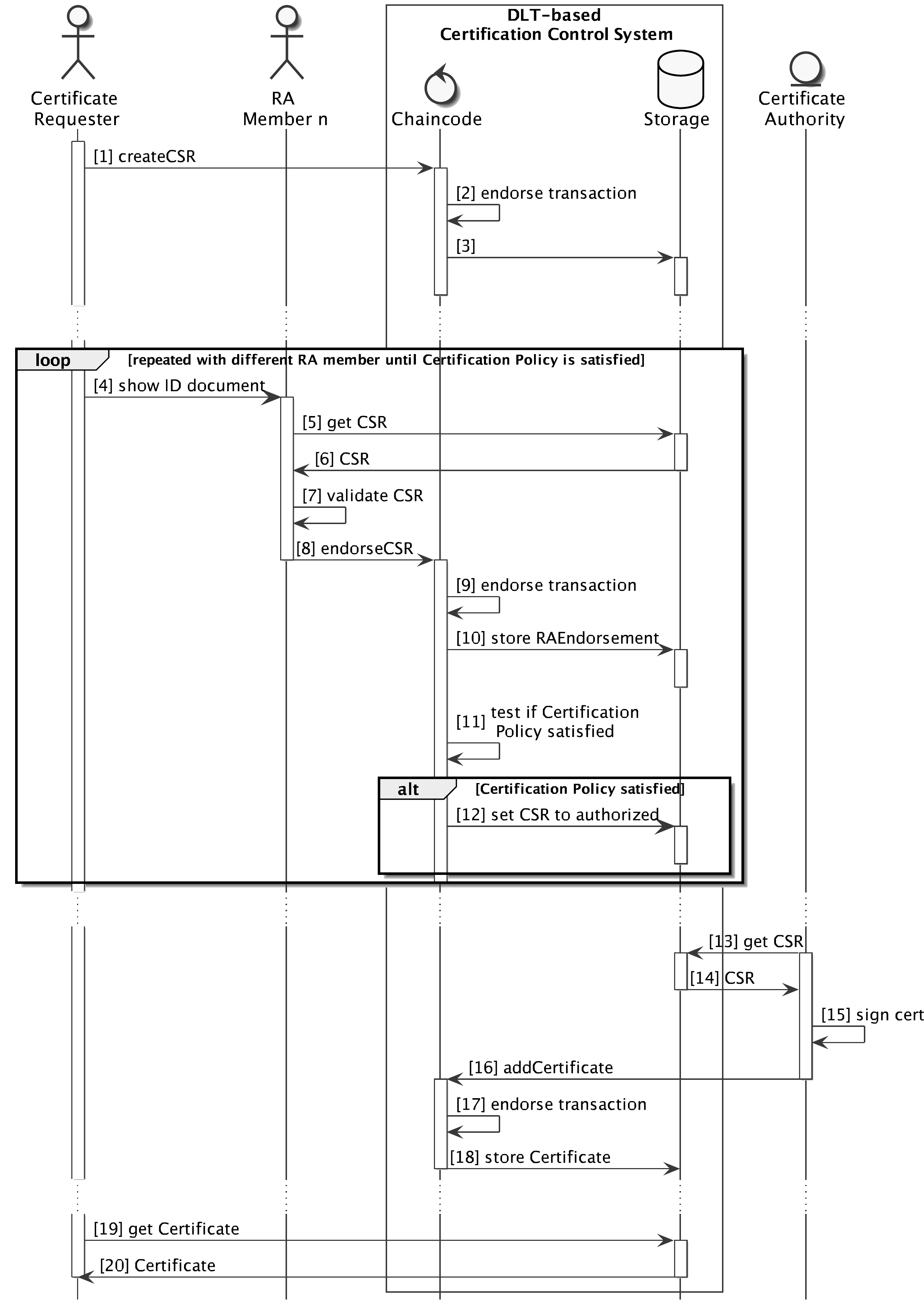}
  \caption{Interaction of entities and processing of user input (simplified)}
  \label{process}
\end{figure}

A new CSR is submitted for validation to the CCS by a Certificate Requester using a user-specific application.
The app sends a \texttt{createCSR} transaction to the appropriate chaincode running on the Fabric network whose purpose it is to add a new \texttt{CSR} record to the \texttt{User} record related to the certificate requester.
Before this transaction can be executed it needs to be endorsed by chaincode, see steps 1--3 in Figure~\ref{process}.

A \texttt{CSR} record contains a unique ID (\texttt{CSRID}), a list of \texttt{RAEndorsement} records, and finally the \texttt{authorized} flag.
This flag indicates the status of a CSR and is initially set to \texttt{false} to indicate that the authorization process is incomplete.

When the user meets with an RA member, the RA member retrieves the \texttt{User} record of that user that contains the CSR from the ledger using an RA-specific app.
Now, the RA member performs various validation steps as discussed in Section~\ref{sec:back:pki} and creates a \texttt{RAEndorsement} record for documentation purposes.
Finally, the record is signed by the RA member to prove authenticity, see steps 5--7.

To add the \texttt{RAEndorsement} record to the related \texttt{CSR} record, the RA member's app sends an \texttt{endorseCSR} transaction into the Fabric network.
Again, chaincode must endorse the transaction, i.e., check if this RA member is authorized to validate this CSR.
To do this, the chaincode retrieves the permissions of the RA member from the data structures that model the certification policy.
The chaincode also validates the signature of the \texttt{RAEndorsement} record and finally adds the \texttt{RAEndorsement} record to the \texttt{CSR} record, see steps 8--10.

After processing the \texttt{endorseCSR} transaction, the chaincode also checks whether the conditions are already met to regard the CSR as being authorized, i.e. if a sufficient number of RA members have successfully validated the CSR.
Optionally, the chaincode can also perform plausibility checks for data contained in the different \texttt{RAEndorsement} records in the last step.
Such checks can be used, for example, to verify that the data about the certificate requester provided by different RA members is identical.
If all checks are successful, the chaincode sets the \texttt{authorized} flag of the \texttt{CSR} record to \texttt{true}, which concludes the verification and authorization process of this CSR, see steps 11 and 12.
Otherwise, the certificate requester must meet with another RA member.

Finally, the CA retrieves the now authorized \texttt{CSR} record from the ledger and issues the certificate. The certificate is added as a \texttt{Certificate} record to the \texttt{User} record and retrieved by the certificate requester's app, see Steps 13--20.

 \section{Discussion}
\label{sec:discuss}

This section discusses whether our system meets the requirements defined in Section~\ref{sec:analysis:reqs},
how well the attack vectors from Section~\ref{sec:analysis:threat} are mitigated,
and further properties.
Table~\ref{tab:coverage} gives an overview which attack vector is covered by which required system property.

If specified by the Certification Policy, the verification of a CSR is performed by multiple RA members.
This approach offers improved protection against mistakes and fraud (R1).
The exact number of how many compromised or malicious RA members, or how many flaws of RA members can be tolerated, depends on the Certification Policy the system uses.
As a rule of thumb, the more costs are invested in the form of trained personnel involved in the CSR validation, the more tolerant the system becomes to failures.
As a failure, we define the issuance of an incorrect or fraudulent certificate.

Formula~\ref{eq1} quantifies the probability of a failure more precisely using simple combinatorics. $n$ denotes the total number of RA members, $m$ the number of malicious RA members, and $v$ is the number of \emph{unanimous} and successful CSR validations required by the system.

\begin{equation}
  \label{eq1}
  P_{fail} = \binom{m}{v}/\binom{n}{v}
\end{equation}

To provide a numerical example, let us assume that $n = 10$ RA members exist, that $m = 3$ of them are malicious, and that $v = 3$ unanimous and successful CSR validations are required by the certification policy.
In this case, the probability of a failure $P_{fail}$ is about 1\%.
Obviously, if more unanimous and successful CSR validations are requested by the system as malicious RA members exist, no failures can occur anymore (e.g. $n = 10$, $m = 3$, $v = 4$, $P_{fail} = 0$), as not enough malicious RA members exist to support the fraudulent CSR.

In a scenario with an increased number of malicious or compromised RA members, $P_{fail}$ will inevitably rise.
However, if an incorrect or fraudulent certificate is created and discovered later, metadata collected by our system will allow the identification of RA members involved in its issuance.
If it is determined that an RA member is repeatedly involved in the issuance of incorrect/fraudulent certificates, she can be retrained or removed from the system to prevent further harm.

Our system is implemented in chaincode running on a Fabric network (R2).
Individual nodes, on which chaincode instances run, are of course capable of cheating.
However, as long as there are honest nodes in the network, attacks on the CCS involve cheating Fabric nodes can be tolerated similarly to cheating RA members.

Requirements R3 and R4 (accountability of CSR validation and certificate issuance) are met because actions of all involved entities require transactions that update the distributed ledger with respective metadata.
This logged metadata is also strongly protected concerning integrity and availability (R5), as the distributed Fabric network replicates the data and protects it against unauthorized modification.

So far, we argued that our system satisfies R1--R5 and protects against failures resulting from AV2--AV4.
However, we have to note that our system is inherently unable to prevent failures caused by direct attacks on the CA that, e.g., result in the attacker having access to the CA's private key (AV1).
However, for certificates issued directly, no validation history that consists of data elements signed by RA members is stored in the distributed ledger.
If the correctness of a certificate in in question, the ledger can be queried comparable to a Certificate Transparency~\cite{rfc6962} log, also see Section~\ref{sec:rw}.
If it is determined that there is no data about the certificate, it can be revoked in further steps.

\begin{table}[]
\centering
\caption{Attack vector coverage by system properties}
\label{tab:coverage}
\begin{tabular}{l|cccc}
& AV1 & AV2 & AV3 & AV4 \tabularnewline
\hline
R1 & & \newmoon & \newmoon & \newmoon \tabularnewline
R2 & & \newmoon & \newmoon & \newmoon \tabularnewline
R3 & & \newmoon & \newmoon & \newmoon \tabularnewline
R4 & \newmoon & & &\tabularnewline
R5 & \newmoon & \newmoon & \newmoon &\tabularnewline
\end{tabular}
\end{table}

Performance of a system influences user acceptance.
In a previous paper~\cite{Geyer2019netsys}, we presented a performance evaluation of Fabric, which indicates that, depending on network size, topology and other parameters, Fabric will be able to cope with the workload of our system without causing a noticeable additional delay in the certification process.
However, several steps of the certification process involve the input of humans and are intentionally designed to be redundant for security reasons.
This approach clearly leads to a high level of commitment on the part of the certificate requesters and the RA.
For this reason, a good balance between security and user acceptance needs to be found.

The last interesting property we want to discuss results from using Fabric: In a setting comparable to the DFN-PKI, the CCS can be distributed across the stakeholders that use it, e.g. different universities.
This distribution is advantageous because it avoids a centralized organization that its members have to trust.
For this reason, the robustness of the resulting Fabric network and thereby of our system is increased as it becomes more difficult to manipulate a considerable amount of Fabric nodes that execute chaincode.
 \section{Related Work}
\label{sec:rw}

This section analyzes related work in the field and compares it to ours.
We begin with standardized additions to the X.509 landscape and continue with alternative approaches.

\emph{Certificate Transparency (CT)}~\cite{rfc6962} provides a public log of certificates.
The log entries are added either by CAs after issuing a new certificate or by clients (like Web browsers) that have received a certificate from a (Web) server.

Clients that want to validate a certificate presented by a server can query the CT log.
If the log contains a different certificate, the client has a strong indicator that there is an issue with the received certificate.

CT is an after-the-fact solution since it cannot prevent a CA from issuing an incorrect certificate.
In that respect, our solution differs from CT as it can prevent the misissuance of certificates in certain cases as discussed earlier.
If a CA has been compromised and the attacker had direct access to the CA's signing key, our solution offers a service comparable to CT, since clients validating a certificate can query the distributed ledger of our system comparable to a CT log.

\emph{Instant Karma PKI (IKP)}~\cite{ikp} is an approach that follows the idea to incentivize diligence and correct behavior of X.509 CAs.
Using Ethereum~\cite{ethereum2018} smart contracts, CAs deposit money in the form of the Ethereum cryptocurrency and agree on paying a penalty in case they misissued a certificate to an illegitimate entity.
This agreed penalty is split and paid out to the owner of the identity affected and to the entity that found the incorrect certificate and reported it to IKP.

IKP and our work share similar goals, i.e., to prevent certificates from being  misissued.
However, the way how this objective is realized differs.
We take a more technical approach that hardens various aspects of certificate issuance.
IKP creates a stimulus for CAs to increase security but does not answer the question of how this goal can be achieved.

\emph{uPort}~\cite{uPort} and \emph{Sovrin}~\cite{sovreign} are solutions for self-sovereign ID management that leverage a web-of-trust approach.
uPort allows users to assign signed pieces of information (so called \emph{credentials}) to their own or a foreign digital identity.
A credential, for instance, can be information about the user, such as her name, address, age, or public key.

Credentials can be made accessible by the user to applications or organizations, for instance, to provide their name and address as part of a registration process for a service.
uPort stores identities in the Ethereum blockchain~\cite{ethereum2018} and uses smart contracts to manage identities, Sovrin is based on Indy~\cite{indy}, a further distributed ledger-based framework of Hyperledger~\cite{hyperpro}.

Besides sharing the common idea of using blockchains/distributed ledgers as a secure base, our project differs from uPort and Sovrin in that we want to improve the security of X.509-based identity management, uPort and Sovrin are developing an alternative technology.
 \section{Conclusions}
\label{sec:conclusion}
\balance

In this paper, we presented a solution that improves the secure and correct operation of X.509 CAs and RAs by enforcing a policy-defined, multi-party validation and authorization workflow of CSR by involving more than just one RA member.
Besides, our system provides accountability for the CSR validation process for forensic purposes.
Because our system runs on the distributed ledger and smart contract framework Hyperledger Fabric, it is tamper-resistant, both in terms of the integrity of the computer processes that control certificate issuance and the integrity of collected metadata.

Compared to related work, our work helps to actively prevent the misissuance of certificates in case RA members are malicious or compromised.
Besides, our system is also useful as an after-the-fact solution if the CA itself got compromised or too many RA members are fraudulent.
In such cases, collected metadata helps to identify fraudulent certificates or malicious RA members.

In future work, we plan to refine our prototype and increase its modularity to allow flexible adaption of the system to different scenarios, like IoT environments.
Furthermore, we are working on a mechanism based on threshold cryptography to decentralize the actual signing of the certificate to reduce the risk of the CA's signing key being easily compromised.
 \section*{Acknowledgments}
The authors would like to acknowledge the valuable contributions of Yannick Gehring, Felix Hoops, and Julian Roos.

\IEEEtriggeratref{0}
\bibliographystyle{IEEEtran}
\bibliography{ref.bib,rfc.bib}

\end{document}